\documentclass[conference]{IEEEtran}
\IEEEoverridecommandlockouts
\usepackage{cite}
\usepackage{amsmath,amssymb,amsfonts}
\usepackage{algorithmic}
\usepackage{graphicx}
\usepackage{textcomp}
\usepackage{xcolor}
\def\BibTeX{{\rm B\kern-.05em{\sc i\kern-.025em b}\kern-.08em
    T\kern-.1667em\lower.7ex\hbox{E}\kern-.125emX}}
\begin{document}

\title{Pattern-based Long Short-term Memory for Mid-term Electrical Load Forecasting\thanks{The project financed under the program of the Polish Minister of Science and Higher Education under the name "Regional Initiative of Excellence" in the years 2019 - 2022 project number 020/RID/2018/19, the amount of financing 12,000,000.00 PLN.}\\
}

\author{\IEEEauthorblockN{Pawe\l { }Pe\l ka and Grzegorz Dudek}
\IEEEauthorblockA{\textit{Department of Electrical Engineering} \\
\textit{Czestochowa University of Technology}\\
Częstochowa, Poland \\
p.pelka@el.pcz.czest.pl, dudek@el.pcz.czest.pl}
}

\maketitle

\begin{abstract}
This work presents a Long Short-Term Memory (LSTM) network for forecasting a monthly electricity demand time series with a one-year horizon. The novelty of this work is the use of pattern representation of the seasonal time series as an alternative to decomposition. Pattern representation simplifies the complex nonlinear and nonstationary time series, filtering out the trend and equalizing variance. Two types of patterns are defined: x-pattern and y-pattern. The former requires additional forecasting for the coding variables. The latter determines the coding variables from the process history. A hybrid approach based on x-patterns turned out to be more accurate than the standard LSTM approach based on a raw time series. In this combined approach an x-pattern is forecasted using a sequence-to-sequence LSTM network and the coding variables are forecasted using exponential smoothing. A simulation study performed on the monthly electricity demand time series for 35 European countries confirmed the high performance of the proposed model and its competitiveness to classical models such as ARIMA and exponential smoothing as well as the MLP neural network model.     
\end{abstract}


\section{Introduction}
The power system load is characterized as a nonlinear and nonstationary process that can undergo rapid changes due to
several factors such as weather, variability of seasons, macroeconomic variations, electricity prices, geographical conditions, and consumer types and their habits. Due to the lack of large-scale energy storage technologies available, the power system should ensure a supply of electricity at any time to cover current demand. Therefore, load or electricity demand forecasting is an essential tool for power system operation and planning. Mid-term electrical load forecasting (MTLF) involves forecasting the daily peak load for following months as well as monthly electricity demand. The latter problem is the subject of this work. 

MTLF is necessary for maintenance scheduling, hydro-thermal coordination, planing of fuel reserve and energy import/export,  and also security assessment. Deregulated power systems need MTLF to be able to negotiate forward contracts. Therefore, the forecast accuracy translates directly into financial performance for energy market participants. All these important reasons explain why new forecasting tools for MLTF are still being developed. They can be roughly divided into a conditional modeling approach and an autonomous modeling approach \cite{Ghi06}. The former approach focuses on economic analysis and long-term planning and forecasting of energy policy. Many input variables are taken into account describing socio-economic conditions, population migrations and power system and network infrastructure. Examples of the conditional modeling approach can be found in \cite{Kan02}, \cite{Bun09} and \cite{Moh18}. In the autonomous modeling approach, the input variables include only historical loads or, additionally, weather factors \cite{Dov99}, \cite{Pei11}, \cite{Pel19}, \cite{Pel19b}.

MTLF forecasting models are built using classical statistical/econometrics tools or machine learning tools \cite{Sug11}. The former include ARIMA, exponential smoothing (EST) and linear regression. ARIMA and ETS can deal with seasonal time series but linear regression requires additional operations such as decomposition or extension of the model with periodic components \cite{Bar01}. Problems with adaptability and nonlinear modeling of the statistical methods has increased researchers' interest in machine learning and AI tools \cite{Gon08}. Of these, neural networks (NNs) are the most popular because of their attractive features including learning capabilities, universal approximation property, nonlinear modeling and massive parallelism. Some examples of using NNs for MLTF are: \cite{Chen17} where NN uses historical loads and weather variables to predict monthly loads and is trained by heuristic algorithms to improve performance, \cite{Gav01} where Kohonen NN is used, \cite{Dov99} where NNs are supported by fuzzy logic, \cite{Pel19} where generalized regression NN is used, and \cite{Pei11} where weighted evolving fuzzy NN is used.

A separate category of NNs are recurrent NNs with connections between nodes forming a directed graph along a temporal sequence. They can exhibit temporal dynamic behavior using their internal state (memory) to process sequences of inputs. Recent works reported that recurrent NNs such as the Long Short-Term Memory (LSTM) NN provide high accuracy on forecasting and outperforms most of the traditional statistical and machine learning methods such as ARIMA, support vector machine and shallow NNs \cite{Yan18}. This is thought to be due to the extra neighboring time frame states dependencies introduced by memory gates.
There are many examples of the application of LSTMs to load forecasting: \cite{Bed18}, \cite{Zhe17}, \cite{Nar17}.

To improve forecasting performance, LSTM is also mixed with other models such as ETS. Such a model, which utilized 100,000 real-life time series, and incorporates all major forecasting methods, including those based on AI and machine learning, as well as traditional statistical ones, won the M4 forecasting competition in 2019 \cite{Mar20}. The winning model developed by Smyl \cite{Smy20} mixes ETS with advanced LSTM. ETS enables the model to capture the main components of the individual time series, such as seasonality and level, while LSTM networks allow nonlinear trends and cross-learning (i.e. using many series to train a single model).    

Taking into the account attractive features of LSTM, we propose in this work pattern-based LSTM forecastig models. The novelty of these models is time series preprocessing by defining yearly patterns. Using a time series composed of patterns of successive years instead of the original time series we simplify the forecasting problem by removing the trend and stabilizing time series variance. So, the forecasting model solves a simplified problem with what we expect will be higher accuracy. After forecasts of the pattern for the next year are generated by the model, we introduce the current trend and variance of the time series to obtain the electricity demand forecasts. We define two types of patterns using current or lagged variables describing the process behavior. The first variant needs these variables to be forecasted for the next year. We use ETS for this.       

The rest of the work is organized as follows. Section 2 describes the electricity demand time series and their representation using yearly patterns. Section 3 gives implementation details of the LSTM forecasting model. Section 4 describes the experimental framework used to evaluate the performance of the proposed models. Finally, Section 5 concludes the work.

\section{Time Series and Their Representations}

\begin{figure}
	\centering
	\includegraphics[width=0.49\textwidth]{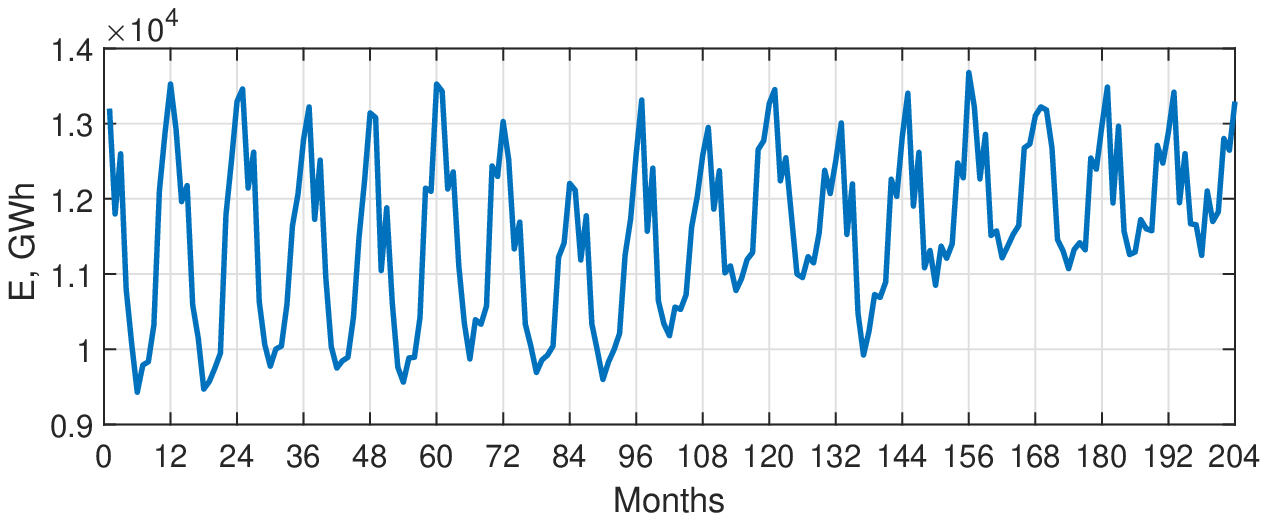}
	\includegraphics[width=0.49\textwidth]{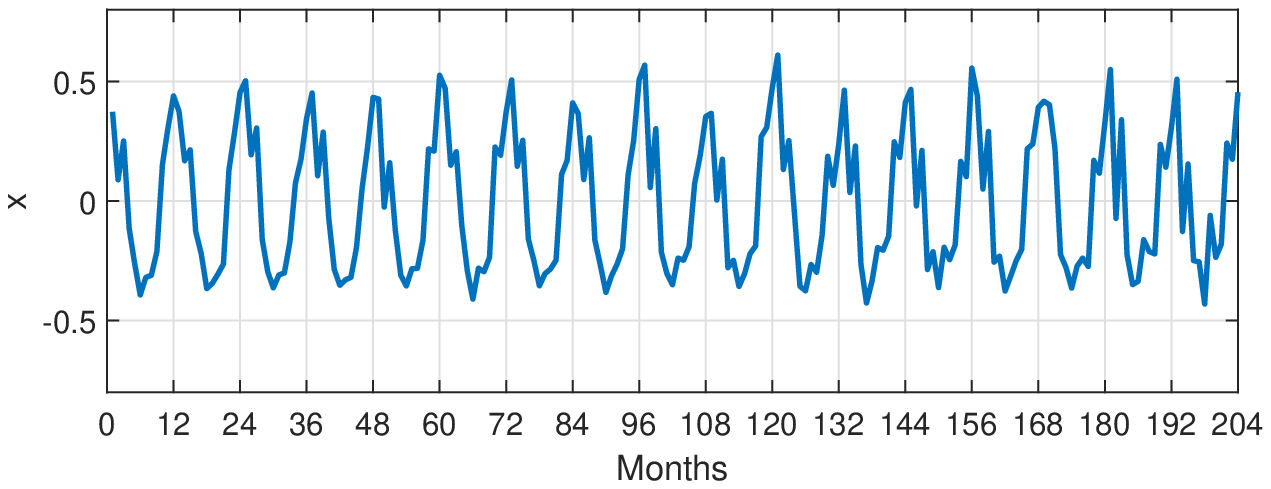}
	\includegraphics[width=0.49\textwidth]{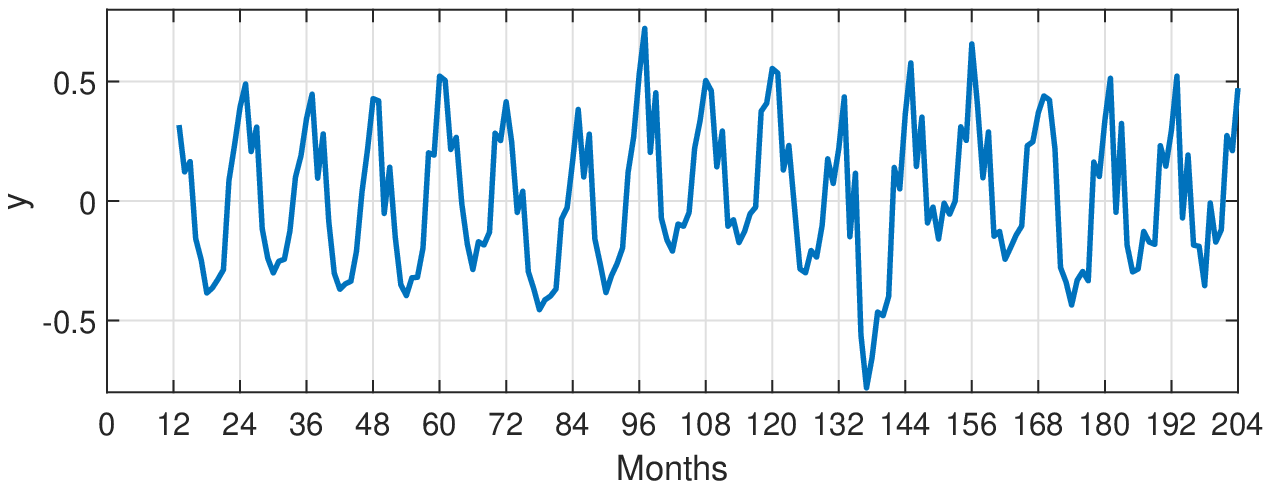}
	\caption{Monthly electricity demand time series for Poland and its x and y representations.} 
	\label{figTSxy}
\end{figure}

Monthly electricity demand time series usually express a trend, yearly cycles and random component. The upper panel of Fig. \ref{figTSxy} depicts an example of such a time series for Poland in the period 1998-2014. Note the nonlinear trend and strong yearly cycles with changing pattern over the years. Additionally, the standard deviation of the yearly cycles changes significantly over time: from $1483$ MWh in 1998 to $696$ MWh in 2014.  

The time series representation is a key component in the construction of forecasting models. The goal is to simplify the forecasting problem and relationships between forecasted variables and predictors. The simplified problem can be solved using simpler models which produce more accurate forecasts. A typical approach for time series preprocessing is to decompose it into trend, seasonal and irregular components. After decomposition, the components showing less complexity than the original time series can be predicted using simpler models. A versatile and robust method for decomposing time series often used in practice is STL (seasonal and trend decomposition using Loess) \cite{Oli18}. Another popular method for seasonal time series decomposition is a wavelet transform \cite{Ben06}. This approach produces the local representation of the time series in both time and frequency domains. Yet another approach to deal with the complex nonlinear and nonstationary time series is Empirical Mode Decomposition \cite{Zhe17} which decomposes a series into so-called intrinsic mode functions without leaving the time domain. 
  
To deal with multiple seasonal cycles and trend in our earlier work, we used similarity-based models operating on patterns of the time series seasonal cycles \cite{Dud15}, \cite{Dud17}. The patterns filter out the trend and those seasonal cycles longer that the basic one and even out variance. They also ensure the unification of input and output variables. Consequently, pattern representation simplifies the forecasting problem and allows us to use models based on pattern similarity. We also built forecasting models operating on patterns using NNs \cite{Pel19}, \cite{Pel19b}, neuro-fuzzy systems, regression trees and other tools. In all cases, pattern representation simplified the problem and led to more accurate forecasts compared to classical models such as ARIMA and ETS. Encouraged by these results, we propose pattern representation for LSTM.

Let us consider the monthly electricity demand time series starting from January and ending in December: $E = \{E_t: t=1,2,...,N\}$. We divide this time series into yearly subsequences $E_i = \{E_t: t=12(i-1)+1, 12(i-1)+2, ..., 12(i-1)+12)\}, i=1, 2, ..., N/12$. Each subsequence can be expressed by a vector $\mathbf{E}_i = [E_{i,1} E_{i,2} … E_{i,12}]^T$. Let us define an x-pattern $\mathbf{x}_i = [x_{i,1} x_{i,2} … x_{i,12}]^T$ as a vector representing a yearly subsequence $E_i$. The function transforming time series elements into patterns depends on the time series character e.g its seasonalities, variance and trend. Some definitions of this function can be found in \cite{Dud15}. In this study we define x-patterns as follows:

\begin{equation}\label{eq4}
x_{i,j} = \frac{E_{i,j}-\overline{E}_i}{\sigma_i}
\end{equation}
where $j = 1, 2, ..., 12$, $\overline{E}_i$ is a mean of sequence $E_i$, and $\sigma_i = \sqrt{\sum_{j=1}^{n} (E_{i,j}-\overline{E}_i)^2}$ is a measure of its dispersion.
 
X-pattern $\mathbf{x}_i$ is a normalized version of vector $\mathbf{E}_i$. Note that yearly subsequences expressed by $\mathbf{E}_i$ have different means and dispersion (see upper panel of Fig. \ref{figTSxy}). After normalization they are unified: all x-patterns have zero mean, the same variance and also a unity of length. They carry information about the shapes of the yearly sequences. Now we create a new time series composed of x-patterns representing successive yearly periods: $x = \{\mathbf{x}_i: i=1,2,...,N/12\} = \{x_{1,1},x_{1,2}, ..., x_{N/12,12}\}$. This time series is shown in the middle panel of Fig. \ref{figTSxy}. Note its regular character and stationarity. 

The forecasting procedure based on x-time series requires the demand forecast based on the x-pattern forecast to be determined. After the x-pattern is generated by the forecasting model, the monthly electricity demands in the forecasted yearly period are calculated from the forecasted x-pattern using transformed equation $\eqref{eq4}$ (this is called decoding). But in this equation the coding variables, $\overline{E}_i$ and $\sigma_i$, are not known, because they are the mean and dispersion of the future yearly subsequence, which has just been forecasted. So, the coding variables should be forecasted from their historical values. We use ETS models for this.

To avoid forecasting the coding variables, we propose another approach. Instead of using the mean and dispersion of the forecasted subsequence as coding variables, we use the mean and dispersion of the preceding subsequence $E_{i-1}$ which is known at the moment of forecasting. In this approach, the yearly subsequence of the time series $E_i$ are represented by a y-pattern $\mathbf{y}_i = [x_{i,1} x_{i,2} … x_{i,12}]^T$ defined as follows:   

\begin{equation}\label{eq5}
y_{i,j} = \frac{E_{i,j}-\overline{E}_{i-1}}{\sigma_{i-1}}
\end{equation}

Although this approach does not guarantee that all y-patterns have the same mean value and variance as in the case of x-patterns, it unifies subsequences of the original time series taking into account the current process variability, expressed by mean $\overline{E}_{i-1}$ and dispersion $\sigma_{i-1}$. The y-patterns representing successive years are composed to create y-time series: $y = \{\mathbf{y}_i: i=2, 3, ...,N/12\} = \{y_{2,1},y_{2,2}, ..., y_{N/12,12}\}$. This time series is shown in the bottom panel of Fig. \ref{figTSxy}. On the basis of the y-time series, the forecasting model learns to predict the y-pattern for the next year. Then, the monthly demands are calculated from transformed equation \eqref{eq5} using known coding variables for the historical sequence $E_{i-1}$.

\section{LSTM Forecasting Model}

The proposed approach can be summarized in the following steps:
\begin{enumerate}
	\item Time series preprocessing.\\
	Depending on a model variant, a raw time series can be used, y-time series or x-time series. In the last case, also the $\overline{E}$-time series and $\sigma$-time series should be defined on the basis of the original time series.   
	\item Time series forecasting.\\
	A sequence-to-sequence regression LSTM model is used in three variants. In the first variant, denoted as LSTM, the model uses raw data to forecast the monthly electricity demand for the next year. In the second variant, LSTMy, the model uses y-time series and forecasts the y-pattern for the next year. In the third variant, LSTMx, the model uses x-time series and forecasts the x-pattern for the next year. In the LSTMx case, also the $\overline{E}$-time series and $\sigma$-time series should be forecatsed for the next year. ETS is used for this.       
	\item Decoding of the forecasted pattern.\\
	The y-pattern predicted by LSTMy is transformed into monthly electricity demand using coding variables determined from the time series history. The x-pattern predicted by LSTMx is transformed using coding variables predicted by ETS.   
\end{enumerate}
The first step was described in detail in Section II. The forecasting LSTM model (step 2) and data post-processing (step 3) are described below.
  
LSTM is a recurrent NN for learning problems related to sequential data \cite{Hoch97}. The main idea behind LSTM is a memory cell which retains its state over time, and non-linear gating units which regulate the information flow in the cell. LSTM is a general model which is very effective at capturing long term temporal relationships and unlike the simple recurrent NNs does not suffer from optimization hurdles, i.e vanishing gradients. In addition to forecasting, the application area of LSTM includes \cite{Gre15}: handwriting recognition, language modeling and translation, acoustic modeling of speech, protein secondary structure prediction, and analysis of audio and video data.      

\begin{figure}
	\centering
	\includegraphics[width=0.4\textwidth]{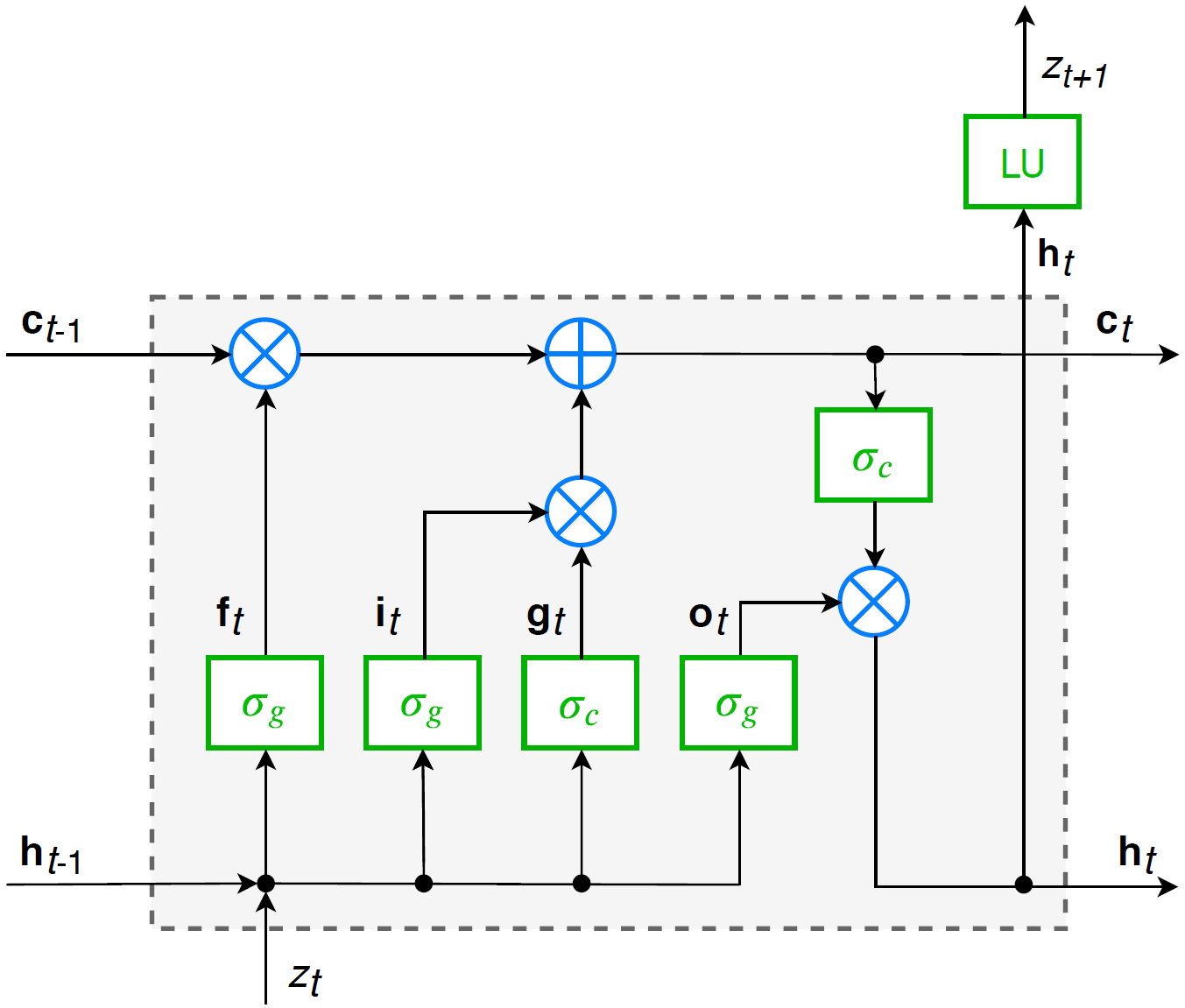}
	\caption{LSTM block.} 
	\label{figLU}
\end{figure}

A diagram of the LSTM block used in this study is shown in Fig. \ref{figLU}. In the diagram, $\textbf{h}_t$ and $\textbf{c}_t$ denote the hidden (or output) state and the cell state at time step $t$, respectively. The cell state contains information learned from the previous time steps. At each time step, information is added to or removed from the cell state. These updates are controlled using three gates: input gate ($i$), forget gate ($f$) and output gate ($o$). At time step $t$, the block uses the current state of the network ($\textbf{c}_{t-1}$, $\textbf{h}_{t-1}$) and the next time step of the sequence ($z_t$) to compute output $\textbf{h}_{t}$ and updated cell state $\textbf{c}_{t}$. Output state $\textbf{h}_{t}$ is further processed by the linear unit (LU) to get the next time series element, $z_{t+1}$, as an output. The hidden and cell states are recurrently connected back to the block input. All of the gates receive the hidden state of the past cycle and the time series element as inputs.

The learnable weights of LSTM are the input weights $\textbf{W}$, the recurrent weights $\textbf{R}$, and the biases $\textbf{b}$. The matrices $\textbf{W}$, $\textbf{R}$ and $\textbf{b}$ are connected with all gates and the cell candidate $g$. Linear unit LU includes the input weights and biases.

The cell state at time step $t$ is given by:
\begin{equation}
\textbf{c}_t = \textbf{f}_t  \otimes  \textbf{c}_{t-1}  + \textbf{i}_t \otimes \textbf{g}_t
\label{eqL1}
\end{equation} 
where operator $\otimes$ denotes the Hadamard product (element-wise product).

The hidden state at time step $t$ is given by:

\begin{equation}
\textbf{h}_t = \textbf{o}_t   \otimes  \sigma_c  (\textbf{c}_t) 
\label{eqL2}
\end{equation} 
where the state activation function $\sigma_c$ is a hyperbolic tangent function.

The following formulas describe the components of LSTM block at time step $t$:

\begin{equation}
\textbf{i}_t= \sigma_g(\textbf{W}_iz_t+\textbf{R}_i\textbf{h}_{t-1}+ \textbf{b}_i)
\label{eqL3}
\end{equation} 
\begin{equation}
\textbf{f}_t= \sigma_g(\textbf{W}_f z_t+\textbf{R}_f\textbf{h}_{t-1}+ \textbf{b}_f)
\label{eqL4}
\end{equation} 
\begin{equation}
\textbf{g}_t= \sigma_c(\textbf{W}_gz_t+\textbf{R}_g\textbf{h}_{t-1}+\textbf{b}_g)
\label{eqL5}
\end{equation} 
\begin{equation}
\textbf{o}_t= \sigma_g(\textbf{W}_oz_t+\textbf{R}_o\textbf{h}_{t-1}+\textbf{b}_o)
\label{eqL6}
\end{equation} 
where gate activation function $\sigma_g$ is a sigmoid function $(1+e^{-x})^{-1}$.

The output forecasted value of the sequence, $z_{t+1}$, is calculated as follows:
\begin{equation}
z_{t+1}= \textbf{W}_z\textbf{h}_{t}+\textbf{b}_z
\label{eqL7}
\end{equation} 

All weights and biases, including: $\textbf{W}_i, \textbf{W}_f, \textbf{W}_g, \textbf{W}_o, \textbf{W}_z \in \mathbb{R}^h$, $\textbf{R}_i, \textbf{R}_f, \textbf{R}_g, \textbf{R}_o \in \mathbb{R}^{h \times h}$, and $\textbf{b}_i, \textbf{b}_f, \textbf{b}_g, \textbf{b}_o, \textbf{b}_z \in \mathbb{R}^h$, are updated based on the difference between the output value $z_{t+1}$ and the actual value
following backpropagation through time algorithm \cite{Wer90}. The number of updated parameters is $4h^2+10h$, where $h$ is a number of hidden units also known as the hidden size. The number of hidden units corresponds to the amount of information remembered between time steps (in the hidden state). The hidden state can contain information from all previous time steps, regardless of the sequence length. If the number of hidden units is too large, then the model might overfit to the training data. This value is adjusted to the time series characteristics and can vary from a few dozen to a few thousand.  

The LSTM forecasting model described above is a sequence-to-sequence regression LSTM network which learns to predict the value of the next time step (the responses are the training sequences with values shifted by one time step). To forecast the values of multiple time steps we predict time steps one at a time and update the network state at each prediction. For each prediction, the previous prediction is used as input to the network. For a better fit and to prevent the training from diverging, we standardize the training data to have zero mean and unit variance. At prediction time, we standardize the test data using the same parameters as the training data.

Depending on the time series representation we use one of the three LSTM models listed below: 
\begin{itemize}
	\item LSTM for forecasting the original time series of the monthly electricity demand $E$,
	\item LSTMx for forecasting x-time series,
	\item LSTMy for forecasting y-time series. 
\end{itemize}

LSTMy and LSTMx generate y-pattern ($\widehat{\textbf{y}}$) or x-pattern ($\widehat{\textbf{x}}$), respectivelly, for the next year. To get the forecasted monthly electricity demands from $\widehat{\textbf{y}}$ we apply transformed equations $\eqref{eq5}$:

\begin{equation}\label{eq9}
\widehat{E}_{i,j} = \widehat{y}_{i,j}\sigma_{i-1} + \overline{E}_{i-1}, \quad j = 1, 2, ..., 12
\end{equation}  
using known coding variables $\overline{E}_{i-1}$ and $\sigma_{i-1}$ for the historical subsequence $E_{i-1}$.

In the case of LSTMx to obtain the forecasted demand values we use transformed equation \eqref{eq4}:
\begin{equation}\label{eq10}
\widehat{E}_{i,j} = \widehat{x}_{i,j} \widehat{\sigma}_{i} + \widehat{\overline{E}}_{i}, \quad j = 1, 2, ..., 12
\end{equation} 

\begin{figure}
	\centering
	\includegraphics[width=0.49\textwidth]{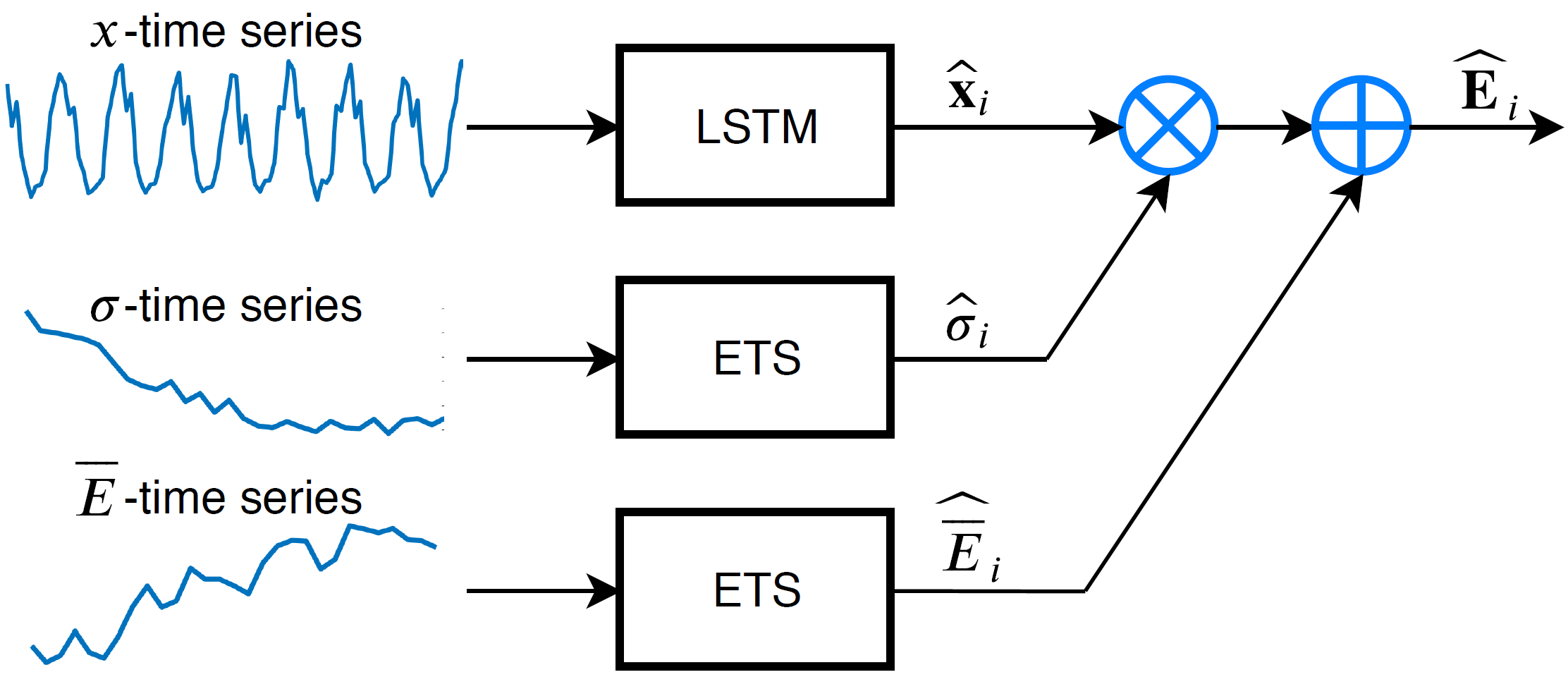}
	\caption{LSTMx model.} 
	\label{figLx}
\end{figure}

The codding variables in this equation, $\widehat{\overline{E}}_{i}$ and $\widehat{\sigma}_{i}$, are not known, because they are the mean and dispersion of the future sequence $E_i$, which has just been forecasted. They are predicted individually from their historical values using ETS. To do so, we prepare two time series: $\overline{E} = \{\overline{E}_i: i=1, 2, ..., N/12 \}$ and $\sigma = \{\sigma_i: i=1, 2, ..., N/12 \}$. Then, we learn two ETS models to generate one step ahead forecasts: $\widehat{\overline{E}}_{i}$ and $\widehat{\sigma}_{i}$. In the next step, we combine the forecasts of $\widehat{x}_{i,j}$, $\widehat{\overline{E}}_{i}$ and $\widehat{\sigma}_{i}$ according to $\eqref{eq10}$.The forecasting model LSTMx architecture is depicted in Fig. \ref{figLx}.      

\section{Simulation Study}
\begin{figure}
	\centering
	\includegraphics[width=0.4\textwidth]{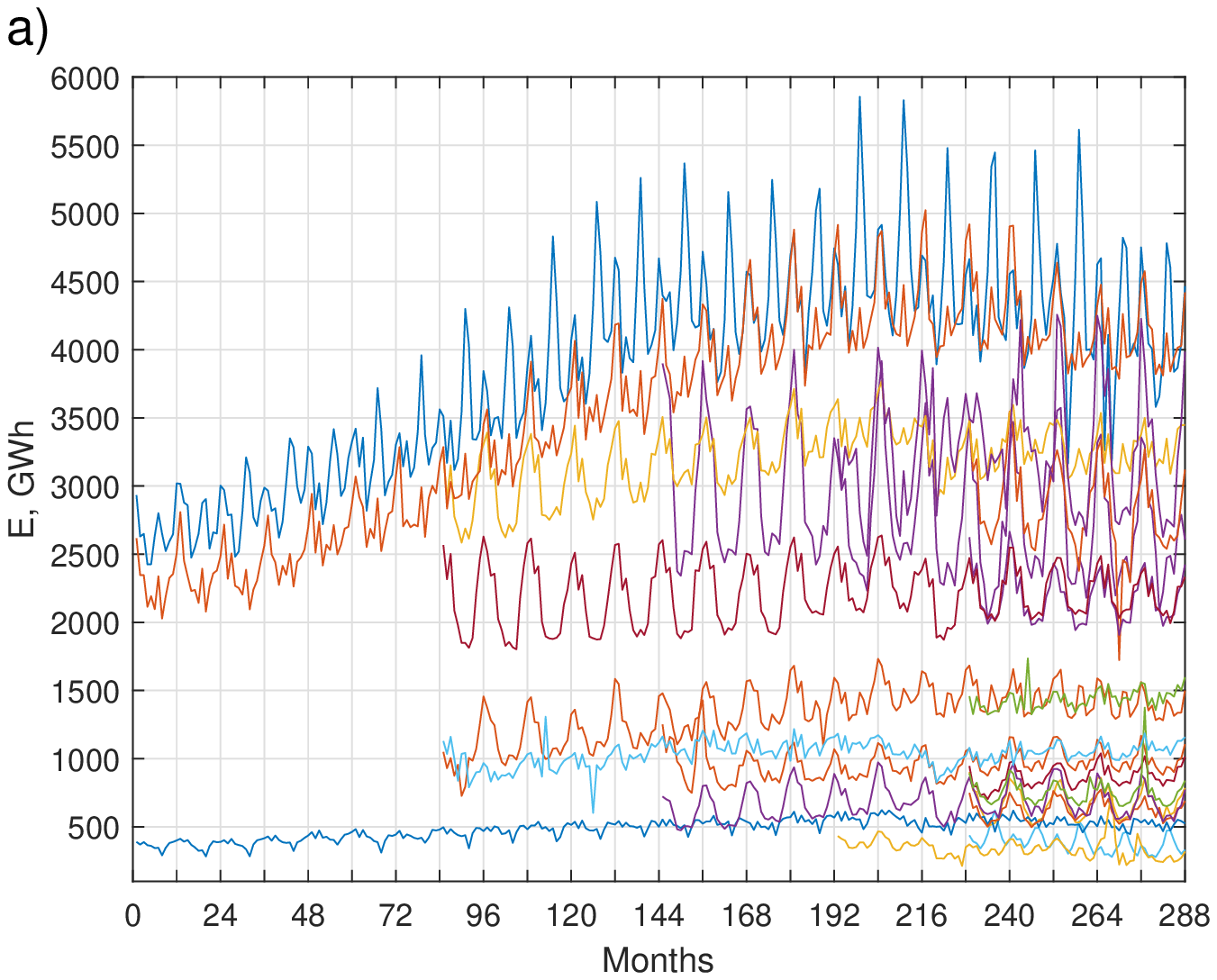}
	\includegraphics[width=0.4\textwidth]{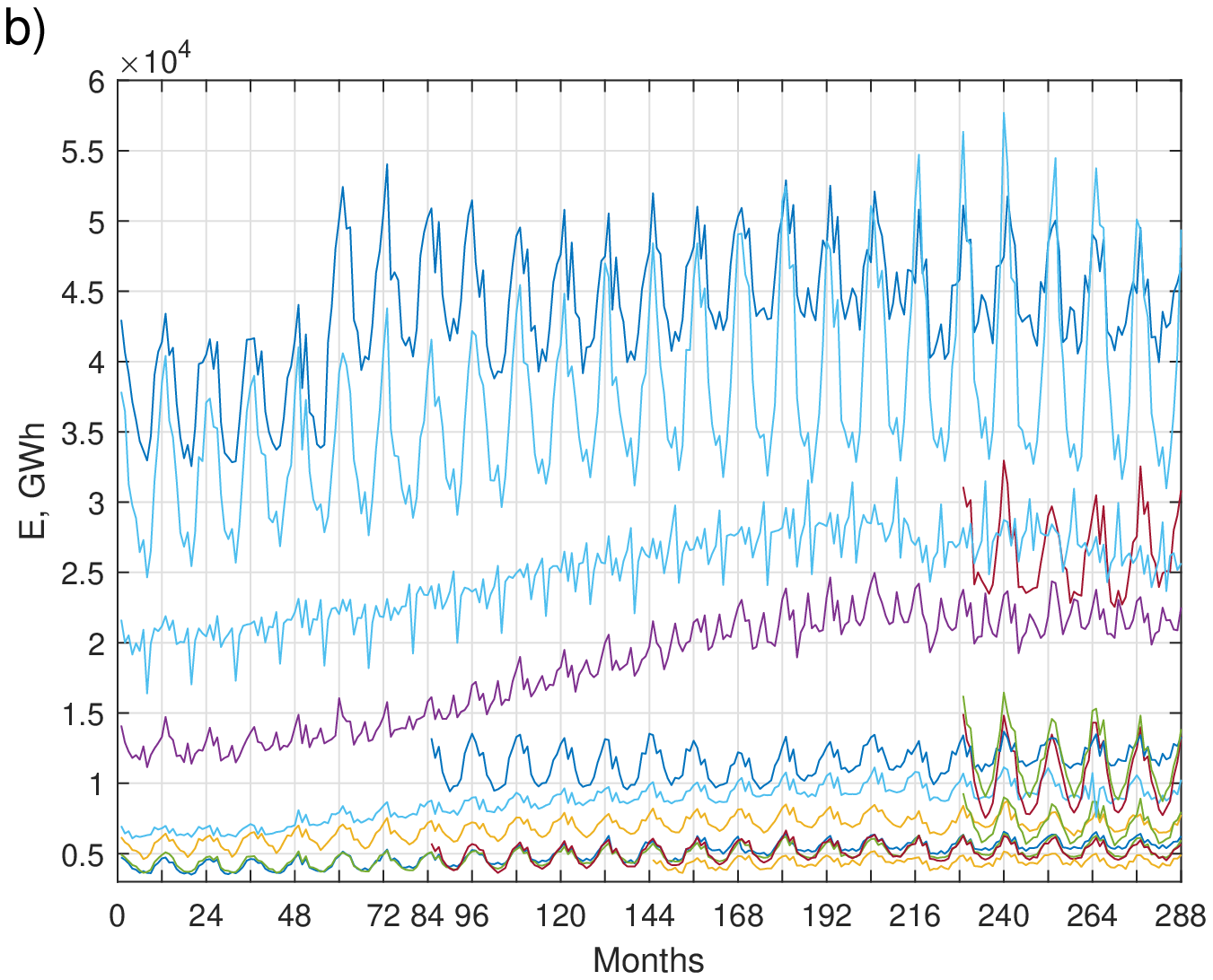}
	\caption{Monthly electricity demand time series for 35 European countries.} 
	\label{figTS}
\end{figure}

The proposed forecasting models based on LSTMs are applied for mid-term load forecasting using real-world data: the monthly electricity demand time series for 35 European countries. The data are taken from the publicly available ENTSO-E repository (www.entsoe.eu). The time series cover different periods: 24 years for 11 countries, 17 years for 6 countries, 12 years for 4 countries, 8 years for 2 countries and 5 years for 12 countries. The models forecast for the twelve months of 2014 (last year of data) using data from the previous period for training. The monthly load time series for 35 countries are shown in Fig. \ref{figTS}. Note that the time series have different lengths, levels, trends, variations and yearly shapes. 

The experiments were carried out using Matlab R2018a implementation of LSTM (function \texttt{trainNetwork} from Neural Network Toolbox). LSTM models were optimized using Adam (adaptive moment estimation) optimizer. The number of hidden nodes was the only hyperparameter which was tuned for each time series. Other hyperparameters remain at their default values: number of epochs - 250, initial learning rate - 0.005. The initial learning rate was dropped after 125 epochs by multiplying by a factor of 0.2. To prevent the gradients from exploding, the gradient threshold was set to 1.

The proposed models were compared with classical statistical models such
as ARIMA and ETS, as well as the neural model (multilayer perceptron, MLP):

\begin{itemize}
\item ARIMA$(p, d, q)(P, D, Q)_{12}$ model implemented in function \texttt{auto.arima} in R environment (package \texttt{forecast}). This function implements automatic ARIMA modeling which combines unit root tests, minimization of the Akaike information criterion (AICc) and maximum likelihood estimation to obtain the optimal ARIMA model \cite{Hyn19}.
\item ETS -- exponential smoothing state space model \cite{Hyn08} implemented in function \texttt{ets} (R package \texttt{forecast}). This implementation includes many types of ETS models depending on how the seasonal, trend and error components are taken into account. They can be expressed additively or multiplicatively, and the trend can be damped or not. As in the case of \texttt{auto.arima}, \texttt{ets} returns the optimal model estimating the model parameters using AICc \cite{Hyn19}.
\item MLP -- multilayer perceptron described in \cite{Pel19b}. This model was designed for MTLF. It learns from patterns defined by \eqref{eq4} and \eqref{eq5}. It predicts one component of y-pattern on the basis of x-patterns. For all 12 components, 12 MLPs are trained and then the forecasts of demands are calculated using \eqref{eq9}. The network has one hidden layer with sigmoidal neurons and learns using Levenberg-Marquardt method with Bayesian regularization to prevent overfitting. The MLP hyperparameters which were adjusted are the number of hidden nodes and length of the input patterns (instead of a fixed value of 12, we select the x-pattern length). We use Matlab R2018a implementation of MLP (function \texttt{feedforwardnet} from Neural Network Toolbox). 
\end{itemize}
We use a single-hidden layer MLP architecture as it has universal approximation capability. Note that when using pattern representation, the relationship between input and output variables is simplified and there is no need to use deeper architectures.

The ETS blocks in LSTMx model (see Fig. \ref{figLx}) were learned using R implementation described above. The LSTM models were  trained 100 times and the final errors were calculated as the averages of errors over 100 independent runs. To asses the dispersion of the forecasts generated by the models in 100 runs we define a dispersion measure as a ratio of standard deviation of forecasts in 100 runs to their median:

\begin{equation}
D = \frac{std(\widehat{E})}{median(\widehat{E})}\cdot 100
\label{eqD}
\end{equation} 
where $\widehat{E}$ is a forecasted energy value.  

\begin{figure}[htbp]
	\centerline{\includegraphics[width=0.49\textwidth]{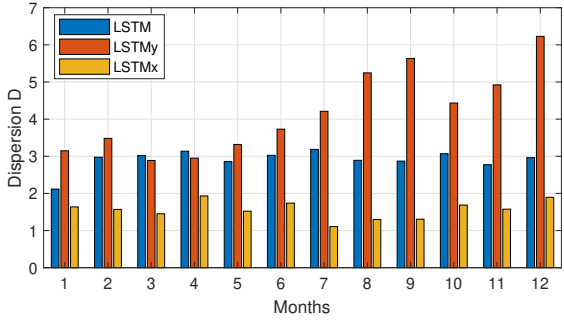}}
	\caption{Dispersion of forecasts for LSTM models.}
	\label{figD}
\end{figure}

Dispersions $D$ for each month of the forecast period are shown in Fig. \ref{figD}. As can be seen from this figure, LSTMx generated the least dispersed forecasts at around 1-2\%. The basic variant of LSTM gave more scattered forecasts, from about 2 to 3\%. LSTMy generated most scattered forecasts (3-6\%) and their dispersion for longer horizons was higher than for shorter ones. This phenomenon was not observed for LSTM and LSTMx.               

\begin{figure}[htbp]
	\centerline{\includegraphics[width=0.49\textwidth]{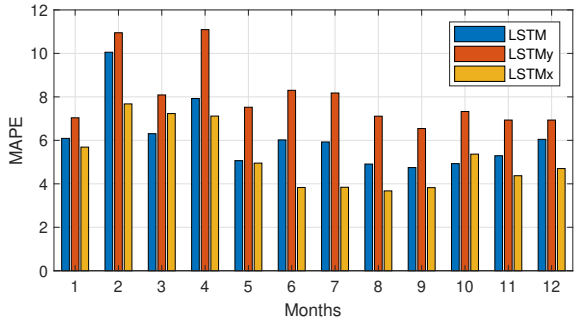}}
	\caption{MAPE for LSTM models.}
	\label{figM1}
\end{figure}

Fig. \ref{figM1} shows the errors for the successive months of the forecasted period (mean absolute percentage errors, MAPEs) averaged over all countries. Note that LSTMx is the most accurate model, while LSTMy is the least accurate one. LSTMy gives the highest errors for each month.    

Table \ref{tabR1} summarizes the accuracy of the models showing median of APE, MAPE, interquartile ranges of APE averaged over all countries and root mean square error (RMSE). LSTMx gives the best results among the LSTM models. This variant may compete with comparative models. It is difficult to select the most accurate model because each error measure indicates a different model as the most accurate: MLP, ETS or LSTMx.  

\begin{table}[htbp]
	\caption{Results comparison among proposed and comparative models.}
	\begin{center}
		\begin{tabular}{|c|c|c|c|c|}
			\hline
			&Median \textit{APE} & \textit{MAPE} & \textit{IQR} &\textit{RMSE}\\
			\hline
			LSTM & 3.73 &	6.11&	4.46&	431.83\\
			\hline
			LSTMy&	3.86&	8.00&	5.67&	500.90\\
			\hline
			LSTMx&	3.08&	5.19&	4.36&	363.22\\
			\hline
			ARIMA&	3.32& 5.65& 5.27& 463.07\\
			\hline
			ETS& 3.50& 5.05& 4.17& 374.52\\
			\hline
			MLP& 2.97& 5.27& 3.89& 378.81\\
			\hline
		\end{tabular}
		\label{tabR1}
	\end{center}
\end{table}

\begin{figure}[htbp]
	\centerline{\includegraphics[width=0.49\textwidth]{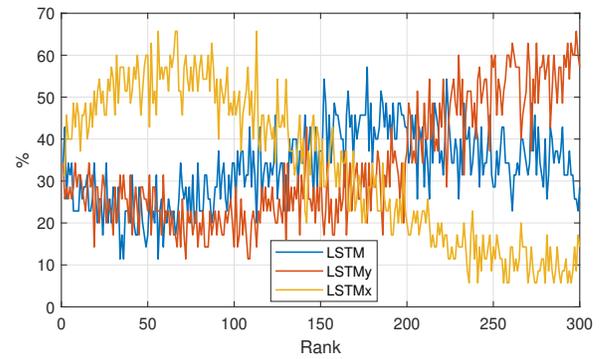}}
	\caption{Ranking of the LSTM models.}
	\label{figR}
\end{figure}

Our experiments involving 100 independent trials for each of 35 countries provide the basis for deeper analysis of the models performances. For each country, we perform a ranking of models taking into account MAPE for each of 100 runs. For our tree models, we have 300 MAPE values which we sort from the smallest to the largest. Then we record the models positions in this ranking, i.e. at each of 300 positions we record the model which took this position. This is repeated for each country. Then we determine how many times a model took $i$-th position. The results expressed as percentages are shown in the Fig. \ref{figR}. As you can see from this figure, LSTMx most often reaches the highest positions in the ranking, and LSTMy most often reaches the lowest positions. 

\begin{figure}[htbp]
	\centerline{\includegraphics[width=0.49\textwidth]{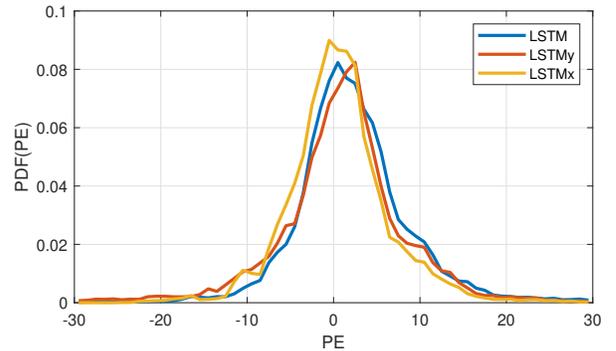}}
	\caption{PDF of the percentage errors.}
	\label{figPDF}
\end{figure}

Fig. \ref{figPDF} depicts empirical probability density functions of the percentage errors (PEs) estimated on the basis of 100 runs and table \ref{tabDS} shows the basic PE descriptive statistics. The PE distributions are similar to the normal one but the tests for the assessment of normality (Jarque–Bera test and Lilliefors test) do not confirm this. In all cases, the forecasts are overestimated, having a positive PE mean. The LSTM model is the most biased and LSTMx the least. Positive values of skewness indicate the right-skewed PE distributions and high kurtosis values indicate leptokurtic distributions where the probability mass is concentrated around the mean.
    
\begin{table}[htbp]
	\caption{Descriptive statistics of percentage errors.}
	\begin{center}
		\begin{tabular}{|c|c|c|c|c|c|}
			\hline
			&Mean & Median & Std & Skewness &Kurtosis\\
			\hline
			LSTM & 3.12&	1.82& 11.79&	5.67&	65.02\\
			\hline
			LSTMy&	2.05&	1.26& 19.33&	6.81&	106.82\\
			\hline
			LSTMx&	1.41&	0.45& 10.61&	5.59&	55.53\\
			\hline
		\end{tabular}
		\label{tabDS}
	\end{center}
\end{table}

\begin{figure}[htbp]
	\centerline{\includegraphics[width=0.49\textwidth]{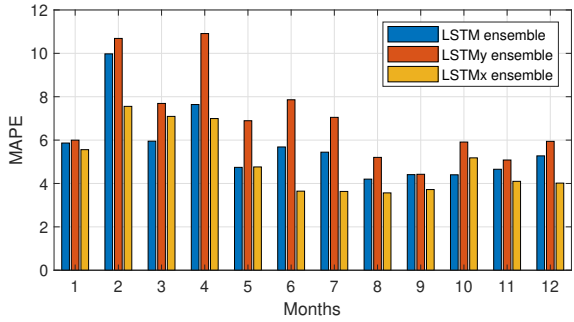}}
	\caption{MAPE for LSTM ensemble models.}
	\label{figEn}
\end{figure}

We can improve accuracy and stability of the learning models by building their ensembles. It was shown that ensembling of
the forecasts enhances the robustness of the method further, mitigating the model and parameter uncertainty
\cite{Pet18}. To built the ensembles we simply combine the forecasts generated in 100 independent trials by averaging (as shown in \cite{Cha18} a simple average of forecasts often outperforms forecasts from single models and a more complicated weighting scheme does not always perform better than a simple average). The results are shown in Table \ref{tabK}. Comparing these to results presented in Table \ref{tabR1}, we can notice a decrease in error for all LSTM models. More detailed results for ensembles are shown in Fig. \ref{figEn}. Comparing to Fig. \ref{figM1}, it can be seen that the greatest improvement in accuracy was for the months of the second half of the year.      

\begin{table}[htbp]
	\caption{Results for LSTM ensembles.}
	\begin{center}
		\begin{tabular}{|c|c|c|c|c|}
			\hline
			&Median \textit{APE} & \textit{MAPE} & \textit{IQR} &\textit{RMSE}\\
			\hline
			LSTM & 3.42&	5.69&	4.81&	318.81\\
			\hline
			LSTMy&	3.43&	6.98&	5.20&	406.37\\
			\hline
			LSTMx&	3.01&	4.99&	4.38&	283.54\\
			\hline
		\end{tabular}
		\label{tabK}
	\end{center}
\end{table}

\begin{figure}[htbp]
	\centerline{\includegraphics[width=0.42\textwidth]{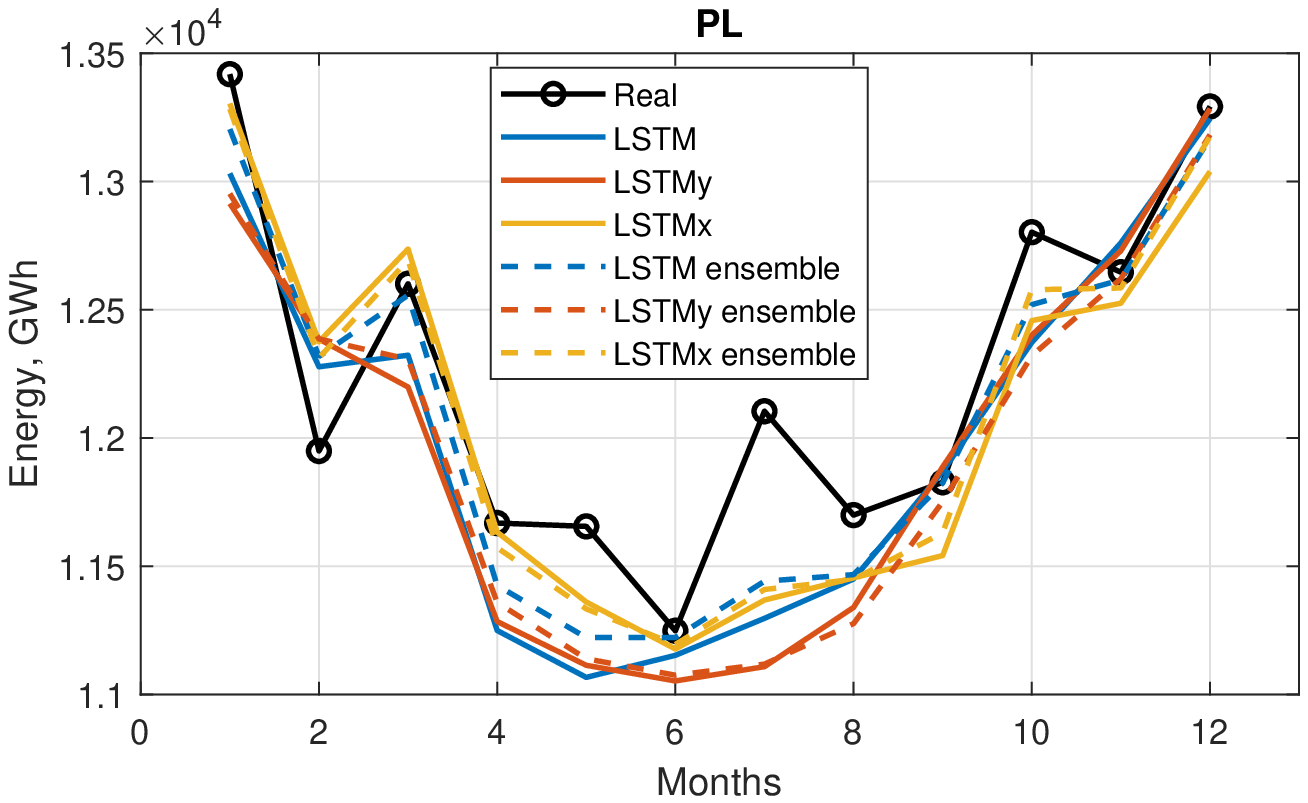}}
	\centerline{\includegraphics[width=0.42\textwidth]{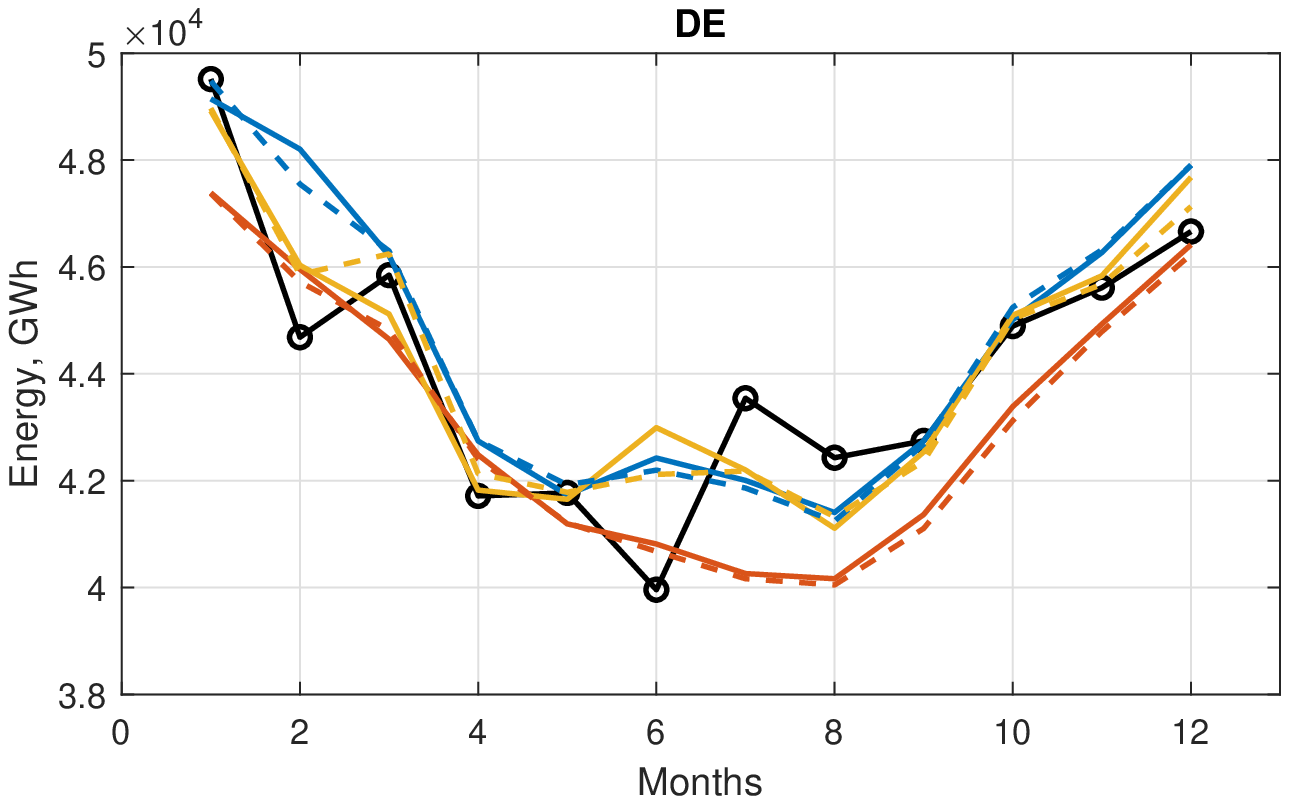}}
	\centerline{\includegraphics[width=0.42\textwidth]{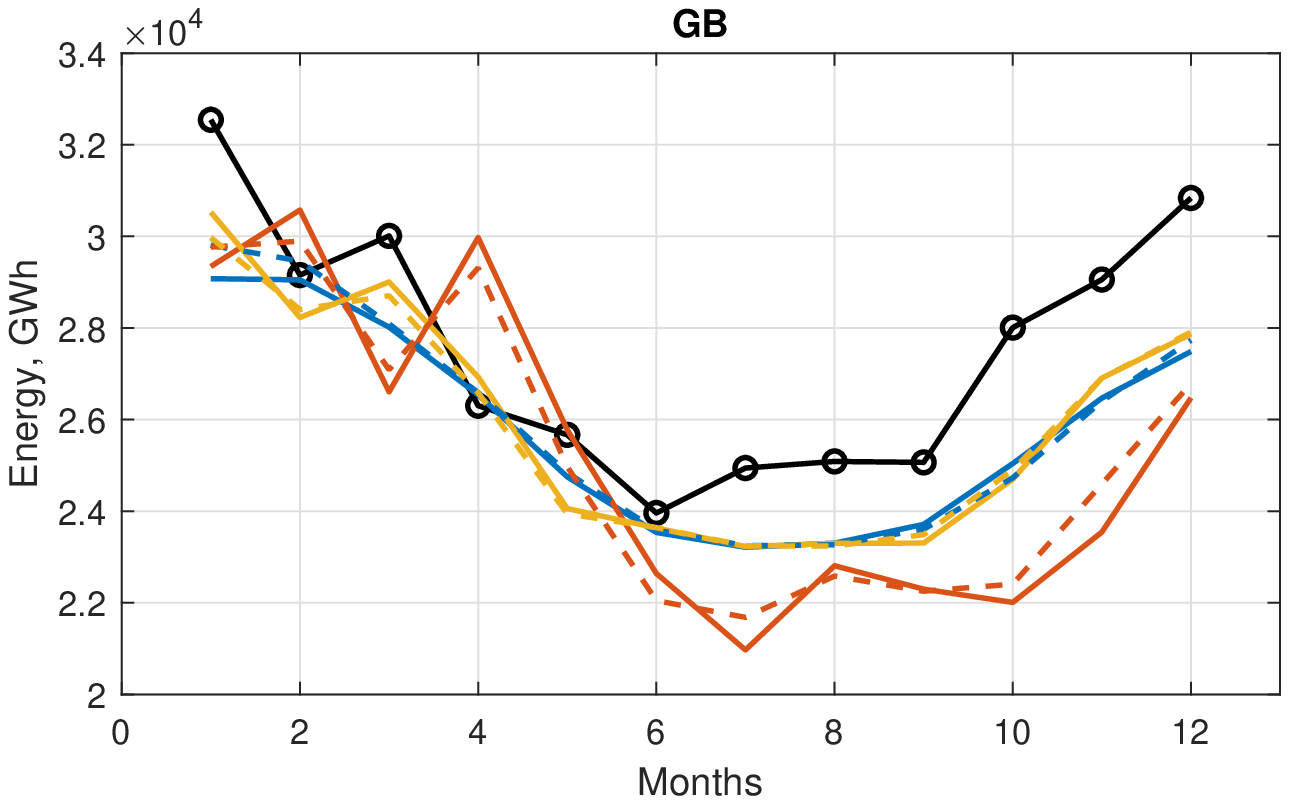}}
	\centerline{\includegraphics[width=0.42\textwidth]{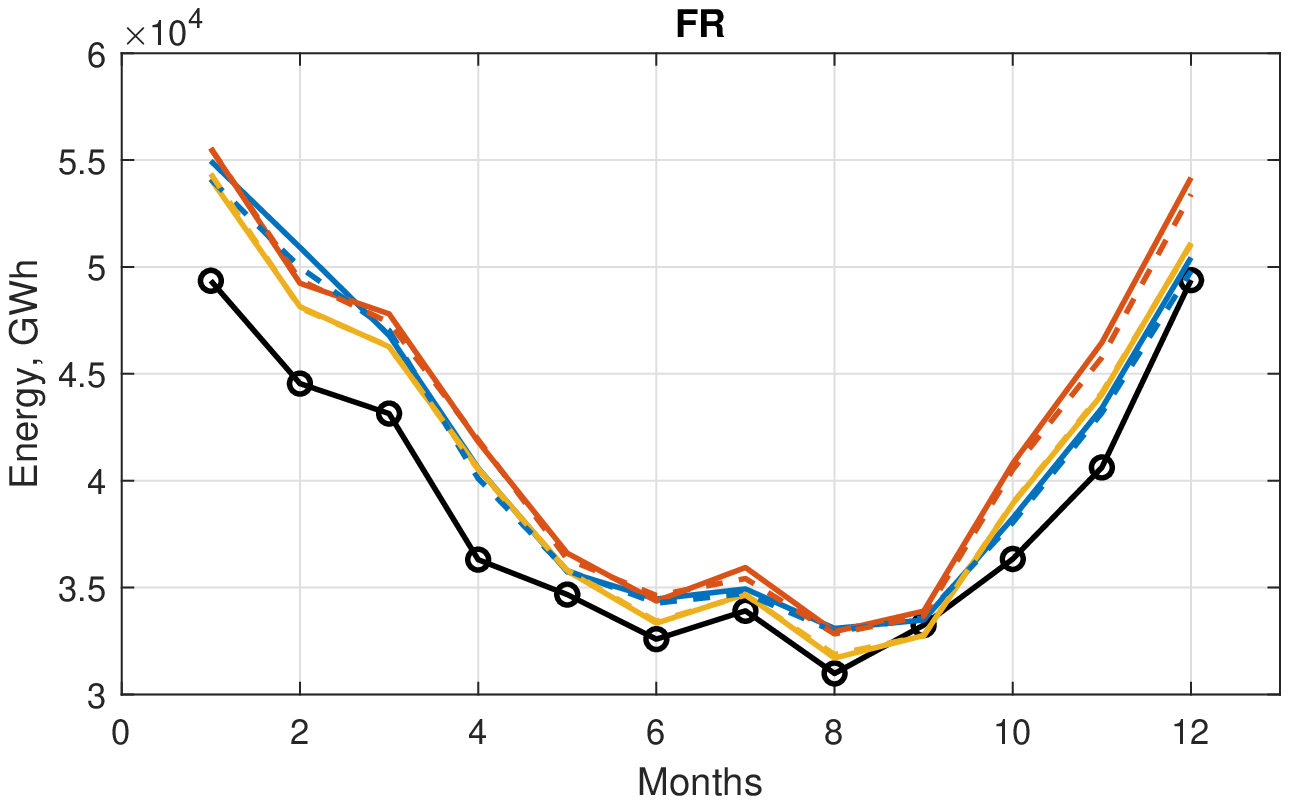}}
	\caption{Examples of the forecasts.}
	\label{figE}
\end{figure}

	Fig. \ref{figE} depicts examples of forecasts generated by the models for four countries. For PL the best forecasts are generated by ensembles of LSTMx ($MAPE=1.79$) and LSTM ($MAPE=1.83$). The individual LSTMx model occupies third place ($MAPE=2.09$). Errors at a similar levels are observed for DE. In this case, the most accurate is LSTMx ($MAPE=1.98$) and its ensemble version ($MAPE=1.60$). For GB the forecasts are underestimated. This results from the fact that demand went up unexpectedly in 2014 despite the downward trend observed in the previous period from 2010 to 2013. The opposite situation for FR caused a slight overestimation of forecasts. For GB errors are around 6\% for LSTM and LSTMx based models and around 11\% for LSTMy based models. LSTMx gives the lowest errors for FR: $MAPE=5.61$ for the individual version and $MAPE=5.71$ for the ensemble version. 

In conclusion of the simulation study, it should be noted that the accuracy of a forecasting model, especially neural model, depends heavily on the appropriate time series preprocessing such as deseasonalization, detrending or decomposition. The LSTM model deals with raw data, without preprocessing, due to recurrent nature and non-linear data processing using gates. However, the proposed x-pattern representation improves LSTM performance. This is because the preprocessed time series is more regular than the original one as it is composed of normalized seasonal subsequences. So, the relationship between input and output variables when using x-pattern approach is simplified and easier to learn. The higher errors for LSTMy are caused by the coding variables, which are determined not from the current seasonal subsequence (represented by the y-pattern), but from the previous one. This disturbs the y-time series (see. Fig. \ref{figTSxy}) and makes it harder to forecast.

In relation to the comparative models, LSTM is much more complex. The number of parameters in LSTM is about 40,000 (assuming the average value of the number of hidden units $h \approx 100$), and many times exceeds the number of parameters in the comparative models. Due to the huge number of parameters and complicated learning procedure using backpropagation through time, the training and optimization time in LSTM is much longer than in comparative models.

\section{Conclusion}

This work proposes pattern-based LSTM forecasting models for mid-term electricity demand forecasting. The key component of the models is the pattern representation which simplifies the complex nonlinear and nonstationary time series, filtering out the trend and equalizing the variance. Two types of patterns are considered: x-pattern and y-pattern. The former requires additional forecasting for the coding variables. The latter determines the coding variables from the process history. Although less complex, the approach based on y-patterns did not achieve the expected improvement in accuracy. Whereas, a combined approach based on x-patterns, LSTMx, turned out to be better than the standard LSTM approach based on a raw time series. In the LSTMx hybrid model, an x-pattern is forecasted using a sequence-to-sequence LSTM network and the coding variables are forecasted using exponential smoothing. 

An experimental study carried out on the electricity demand time series for 35 European countries has shown that LSTMx generates the most accurate forecasts in a twelve-month horizon among LSTM models. The LSTMx results are comparable with those generated by classical statistical models such as exponential smoothing, and the pattern-based neural networks (MLP). Ensembling LSTM models by simple averaging over 100 runs decreases further the forecast errors.


\begin{thebibliography}{00}
	
	\bibitem{Ghi06}
	M. Ghiassi, D. K. Zimbra, and H. Saidane, ``Medium term system load forecasting with a dynamic artificial neural network model,`` Electric Power Systems Research, vol. 76, pp. 302--316, 2006.
	
	\bibitem{Kan02}
	M. S. Kandil, S. M. El-Debeiky, and N. E. Hasanien, ``Long-term load forecasting for fast developing utility using a knowledge-based expert system``, IEEE Trans. Power Syst., vol. 17(2), pp. 491--496, 2002.
	
	\bibitem{Bun09}
	P. Bunnoon, K. Chalermyanont, and C. Limsakul, ``Mid term load forecasting of the country using statistical methodology: Case study in Thailand,`` International Conference on Signal Processing Systems, pp. 924--928, 2009.
	
	\bibitem{Moh18}
	N. A. Mohammed, ``Modelling of unsuppressed electrical demand forecasting in Iraq for long term,`` Energy, vol. 162, pp. 354--363, 2018.
	
	\bibitem{Dov99}
	E. Doveh, P. Feigin, and L. Hyams, ``Experience with FNN models for medium term power demand predictions,`` IEEE Trans. Power Syst., vol. 14(2), pp. 538--546, 1999.
	
	\bibitem{Pei11}
	P. C. Chang, C. Y. Fan, and J. J. Lin, ``Monthly electricity demand forecasting based on a weighted evolving fuzzy neural network approach,``  Electrical Power and Energy Systems, vol. 33, pp. 17--27, 2011.
	
	\bibitem{Pel19}
	P. Pe\l ka and G. Dudek, ``Medium-term electric energy demand forecasting using generalized regression neural network,`` Proc. Conf.  Information Systems Architecture and Technology ISAT 2018, vol. AISC 853, pp. 218--227, Springer, Cham, 2018. 
	
	\bibitem{Pel19b}
	P. Pe\l ka and G. Dudek, ``Pattern-based forecasting monthly electricity demand using multilayer perceptron,`` Proc. Conf. Artificial Intelligence and Soft Computing ICAISC 2019, vol. LNAI 11508, pp. 663--672, Springer, Cham, 2019. 
	
	\bibitem{Sug11}
	L. Suganthi and A. A. Samuel, ``Energy models for demand forecasting — A review,`` Renew Sust Energy Rev, vol. 16(2), pp. 1223--1240, 2002. 
	
	\bibitem{Bar01}
	E. H. Barakat, ``Modeling of nonstationary time-series data. Part II. Dynamic periodic trends,`` Electr Power Energy Systems, vol. 23, pp.  63--68, 2001.
	
	\bibitem{Gon08}
	E. González-Romera, M.A. Jaramillo-Morán, and D. Carmona-Fernández, ``Monthly electric energy demand forecasting with neural networks and Fourier series,`` Energy Conversion and Management, vol. 49, pp. 3135--3142, 2008.
	
	\bibitem{Chen17}
	J. F. Chen, S. K. Lo, and Q. H. Do, ``Forecasting monthly electricity demands: An application of neural networks trained by heuristic algorithms,`` Information, vol. 8(1), 31, 2017.
	
	\bibitem{Gav01}
	M. Gavrilas, I. Ciutea, and C. Tanasa, ``Medium-term load forecasting with artificial neural network models,`` IEEE Conf. Elec. Dist. Pub., vol. 6, 2001.
	
	\bibitem{Yan18}
	K. Yan, X. Wang ,Y. Du ,N. Jin , H. Huang, and H. Zhou, ``Multi-Step Short-Term Power Consumption Forecasting with a Hybrid Deep Learning Strategy,`` Energies, vol. 11(11), 3089, 2018.
	
	\bibitem{Bed18}
	J. Bedi and D. Toshniwal, ``Empirical mode decomposition based deep learning for electricity demand forecasting,`` IEEE Access, vol. 6, pp. 49144--49156, 2018.
	
	\bibitem{Zhe17}
	H. Zheng, J. Yuan, and L. Chen, ``Short-Term Load Forecasting Using EMD-LSTM Neural Networks with a Xgboost Algorithm for Feature Importance Evaluation,`` Energies 2017, vol. 10(8), 1168, 2017.
	
	\bibitem{Nar17}
	A. Narayan and K. W. Hipel, ``Long short term memory networks for short-term electric load forecasting,`` 2017 IEEE International Conference on Systems, Man, and Cybernetics (SMC), Banff, AB, pp. 2573--2578, 2017.
	
	
	\bibitem{Mar20}
	S. Makridakis, E. Spiliotis, and V. Assimakopoulos,
	``The M4 Competition: 100,000 time series and 61 forecasting methods,``
	International Journal of Forecasting,
	vol. 36(1),
	pp. 54--74, 2020.
	
	\bibitem{Smy20}
	S. Smyl, ``A hybrid method of exponential smoothing and recurrent neural networks for time series forecasting,``
	International Journal of Forecasting, vol. 36(1), pp. 75--85, 2020.
	
	
	\bibitem{Oli18}
	E. M. de Oliveira and F. L. C. Oliveira, ``Forecasting mid-long term electric energy consumption through bagging ARIMA and exponential smoothing methods,`` Energy, vol. 144, pp. 776--788, 2018. 
	
	\bibitem{Ben06}
	D. Benaouda, F. Murtagh, J. L. Starck, and O. Renaud, ``Wavelet-based nonlinear multiscale decomposition model for electricity load forecasting,`` Neurocomputing, vol. 70(1–3), pp. 139--154, 2006.
	
	\bibitem{Dud15}
	G. Dudek, ``Pattern similarity-based methods for short-term load forecasting – Part 1: Principles,`` Applied Soft Computing, vol. 37, pp. 277--287, 2015.
	
	\bibitem{Dud17}
	G. Dudek and P. Pe\l ka, ``Medium-term electric energy demand forecasting using Nadaraya-Watson estimator,`` Proc. 18th Int. Scientific Conf. on Electric Power Engineering (EPE'17), pp. 1--6, 2017.
	
	
	\bibitem{Hoch97}
	S. Hochreiter and J. Schmidhuber, ``Long short-term memory,`` Neural Computation, vol. 9(8), pp.1735--1780, 1997.
	
	\bibitem{Gre15}
	K. Greff, R. K. Srivastava, J. Koutnik, B. R. Steunebrink, and
	J. Schmidhuber, ``LSTM: A search space odyssey,`` arXiv preprint
	arXiv:1503.04069, 2015.
	
	\bibitem{Wer90}
	P. J. Werbos,  ``Backpropagation through time: What it does and how to do it,`` Proc. IEEE 1990, vol. 78, pp. 1550--1560, 1990.
	
	
	\bibitem{Hyn19}
	R. J. Hyndman and G. Athanasopoulos, Forecasting: Principles and Practice, 2nd edition, OTexts: Melbourne, Australia. OTexts.com/fpp2 (2018) Accessed on 4 October 2019.
	\bibitem{Hyn08}
	R. J. Hyndman, A. B. Koehler, J. K. Ord, R. D. Snyder, Forecasting with Exponential Smoothing: The State Space Approach, Springer, 2008.
	
	
	\bibitem{Pet18}
	F. Petropoulos, R. J. Hyndman, and C. Bergmeir, ``Exploring
	the sources of uncertainty: Why does bagging for time series
	forecasting work?,`` European Journal of Operational Research, vol. 268(2),
	pp. 545–554, 2008.
	
	\bibitem{Cha18}
	F. Chan and L. L. Pauwels, ``Some theoretical results on forecast
	combinations,`` International Journal of Forecasting, vol. 34(1), pp. 64--74, 2018.
	

	
\end{thebibliography}
\end{document}